\documentclass{PoS}
\usepackage{amsmath,amssymb,bm}
\title{New extended Crewther-type relation and the   consequences 
of   multiloop perturbative results}

\ShortTitle{New extended Crewther-type relation and the  
consequences of multiloop perturbative  results }

\author{\speaker{A.L.Kataev} \\
        Institute for Nuclear Research of the Academy 
of Sciences of Russia, 117312, Moscow, Russia \\
        E-mail: \email{kataev@ms2.inr.ac.ru}}

\author{S.V.Mikhailov 
       \\ Bogoliubov Laboratory of Theoretical Physics,
JINR, 141980, Dubna, Russia \\
       E-mail: \email{mikhs@theor.jinr.ru}}

\abstract{\noindent 
We discuss the current status of the investigations of the conformal symmetry breaking 
contributions to three forms for the QCD generalizations of the Crewther relation. 
The new third form of the extension of this relation is considered in more detail. 
Particular attention is paid to the
discussions of the applications of the $\beta$-expansion formalism proposed previously by one 
of us (SVM).
Several relations between 5-loop contributions to the Adler D-function $D^{NS}_{A}$
and to the polarized Bjorken sum rule $S_\text{Bjp}$ are presented. 
One of them gives the additional confirmation of
the correctness of the advanced analytical computer calculations of order $\alpha_s^4$ 
contributions to $D^{NS}_{A}$ and $S_\text{Bjp}$ in the general $SU(N_c)$ gauge group.         
}
\FullConference{The XIXth International Workshop on High Energy 
Physics and Quantum Field Theory, QFTHEP2010\\
        September 08-15, 2010\\
        Golitsyno, Moscow, Russia}

\begin{document}

\section{Introduction}
There are several complementing each other 
QCD extensions of the original Crewther relation, which was    
derived in \cite{CR}  from the axial-vector-vector (AVV) triangle 
amplitude in the quark-parton model. This fundamental 
relation has the following form: 
\begin{equation} 
\label{Cr}
D \cdot C^{Bjp}  = 1~. 
\end{equation}
The first entry  in the l.h.s of Eq.(\ref{Cr})   
is   defined as the quark-parton model expression for the non-singlet 
part of the $e^+e^-$ Adler function, $D^{NS}_{A}$,
\begin{equation}
\label{DA}
D^{NS}_{A}(a_s)= \left(N_c \sum_f Q_f^2\right) 
\cdot D(a_s) ~  
\end{equation}
while the second term in the l.h.s.  is the quark-parton 
model limit of  
 the non-singlet coefficient 
function  $C^{Bjp}$ of the Bjorken 
sum rule of the polarized lepton-hadron deep-inelastic 
scattering (DIS)
\begin{equation} 
\label{Bjp} 
S_{Bjp}(a_s)= \left(\frac{1}{6}\frac{g_A}{g_V}\right)\cdot 
C^{Bjp}(a_s).  
\end{equation}  
$C^{Bjp}$ also 
enters into the non-singlet part 
of the Ellis-Jaffe  sum rule of 
the polarized lepton-hadron DIS. 

The  derivation of Eq.(\ref{Cr}) is essentially based 
on the concept of the conformal symmetry (CS). It is known 
that CS , which is valid in the quark-parton model limit, 
is the symmetry under the following transformations of coordinates 
(see e.g. the reviews \cite{MMT}, \cite{BKM03})  : 
\begin{enumerate} 
\item 
scale transformation or dilatation  
$x^{'\mu}=\rho x^{\mu}$  with 1 parameter $\rho$>0, 
\item 
special conformal transformations 
$\displaystyle x^{'\mu}=\frac{x^{\mu}+\beta^{\mu}x^2}{1+2\beta x+\beta^2 x^2}$
 with 4 parameters $\beta^{\mu}$ and  
\item 
translations  $x^{'\mu}=x^{\mu}+\alpha^{\mu}$ with 4 parameters $\alpha^{\mu}$, 
\item 
homogeneous Lorentz transformations $x^{'\mu}=\Lambda_{\nu}^{\mu}x^{\nu}$  
that also contain  4 parameters.
\end{enumerate} 
Note that in the case of perturbative quenched QED, namely,  
in the approximation when  QED diagrams  containing   
internal photon vacuum polarization contributions are neglected, 
the original Crewther relation of Eq.(\ref{Cr}) is also valid \cite{ACGJ}.   
Other important quantum field theory  studies  
based on the concept of the CS are described, e.g., 
in  \cite{BKM03}, \cite{Baker:1979zz}, \cite{AGM99}. 

However, it is known that conformal symmetry is broken by the normalization 
of the coupling constant in the renormalized  massless quantum field 
models (for details see, e.g.,  \cite{BSHbook}).
One of the basis results of this procedure is  nonzero normalization group $\beta$--function
that can be defined within perturbation theory. 
The factor   $\beta(a_s)/a_s$ 
  appears as the result of 
renormalization of  the trace of the energy-momentum tensor  
\cite{EllisChanowitz,CR,Nielsen,AdlerCollinsDunkan,
CollinsDunkanJoglekar,PMinkowski} and generates  the conformal anomaly.
Here $a_s=\alpha_s/\pi$, and $\alpha_s$ is the QCD coupling.

The original Crewther relation of Eq.(\ref{Cr}) can be generalized 
to the QCD case in the following way: 
\begin{equation} \label{CSPp}
D_{}(a_s(Q^2))\cdot C^{Bjp}(a_s(Q^2)) = 1 + \Delta_{\rm{csb}}(a_s(Q^2)),
\end{equation}
 where $Q^2$ is the Euclidean  transfered momentum  and 
$\Delta_{\rm{csb}}(a_s(Q^2))$ is the conformal-symmetry breaking 
(CSB) term. 

Three  forms for the QCD   generalizations of the  Crewther relations 
are known at present. 
Within the    first  generalization the additional 
correction to the ``Crewther unity''  in the  $\overline{MS}$-scheme reads 
\begin{equation}
\label{BK}
 \Delta_{\rm{csb}}(a_s)=\bigg(\frac{\beta(a_s)}{a_s}\bigg)
P(a_s)= \bigg(\frac{\beta(a_s)}{a_s}\bigg) \sum_{m\geq 1}  K_{m} a_s^m \,.
\end{equation}
This expression was discovered in \cite{BK93}, where  the 
analytical expressions of  two terms  $K_{i}$ with  $i\leq 2$ was fixed. 
The work \cite{BK93} was based on  careful  inspection of  the $SU(N_c)$ group 
structure of the available (to the moment) perturbative approximations  
for  the  functions  $D(a_s)$ and $C^{Bjp}(a_s)$  
and of the  2-loop expression for the QCD $\beta$-function. 
The NLO corrections to the $D$-function were known  from the  calculations  
performed both analytically \cite{Chetyrkin:1979bj} 
and numerically  \cite{Dine:1979qh} (the analytical results were confirmed in
\cite{Celmaster:1979xr}), 
while the analytical N$^2$LO perturbative corrections were evaluated in \cite{GorKatL},  
\cite{SurgSam} and confirmed later on in \cite{Chetyrkin:1996ez} with the help of 
a different theoretical technique.
Analogous NLO corrections to $C^{Bjp}$   
calculated in \cite{Gorishnii:1985xm} 
were  confirmed later on in \cite{Zijlstra:1992kj}. 
The corresponding  N$^2$LO corrections were  
evaluated in \cite{LarinVermaseren}.   

The validity of the expression (\ref{BK}) in all orders 
of perturbation theory  was explored  in the momentum space  
\cite{GabadadzeKataev} (for some details see \cite{Kataev:1996ce}) 
and proved rigorously  in the coordinate space \cite{Cr97} 
(for  additional discussions see \cite{BKM03}).   
The explicit expression for the  N$^3$LO 
term $K_3$  in  Eq.(\ref{BK}) was obtained  recently  \cite{Baikov:2010je}. 
It was found as the result of multiplication of  $SU(N_c)$  group expression 
 for the ${\cal O}(a_s^4)$ corrections to the   $D(a_s)$\footnote{ 
In the case of $N_c$=3 numbers of colours,  
this correction is  known from analytical calculations of \cite{Baikov:2008jh}.}  
and  $C^{Bjp}(a_s)$ 
functions  the analytically  calculated in \cite{Baikov:2010je}  with  taking into account the  3-loop 
analytical    approximation  of  the QCD $\beta$-function, originally  evaluated   
in  \cite{Tarasov:1980au} and  confirmed later on  in  \cite{Larin:1993tp}.

In the  second generalization of the Crewther relation formulated 
in \cite{Brodsky:1995tb} at N$^2$LO,  
the CSB-term $\Delta_{\rm{csb}}(a_s(Q^2))$ in Eq.(\ref{CSPp}) is absorbed in the 
scale $Q^{*2}_{D}$  of the effective charge of  the $D$-function 
$\hat{a}_{Bjp}(Q^2)=\alpha^{eff}_{s}(Q^2)/\pi$ by using the 
BLM scale-fixing procedure \cite{Brodsky:1982gc} extended firstly    
to the N$^2$LO in \cite{Grunberg:1991ac}. 
As the result,  
the second generalized form of the Crewther relation takes 
the following N$^2$LO form: 
\begin{equation}
\label{BS}
\bigg(1+\hat{a}_{D}(Q^{*2}_{D})\bigg)\bigg(1- \hat{a}_{Bjp}(Q^2)\bigg)=1             
\end{equation}
where $\hat{a}_{D}(Q^2)$ and $\hat{a}_{Bjp}(Q^2)$ are the effective 
charges of the $D(a_s(Q^2))$ and  $C^{Bjp}(Q^2)$ functions. 
The concept of the effective charges was introduced  
in the process of creation of the effective charges approach developed  
in \cite{Grunberg:1980ja}, \cite{Krasnikov:1981rp}, \cite{Grunberg:1982fw}. 

However, the  all-order proof of the  second generalization of the 
Crewther relation of Eq.(\ref{BS}) is absent. 
In view of this  the analysis of \cite{Baikov:2010je}, 
which obviously demonstrated, that Eq.(\ref{BS}) is also  satisfied  at N$^3$LO, 
can be considered as an important step of the verification of the second form of the 
generalization of the Crewther relation in high orders.   

It is also  possible to use the N$^2$LO variant of the  BLM prescription  
\cite{Grunberg:1991ac} to absorb the  N$^2$LO approximation 
of the  CSB-term $\Delta_{\rm{csb}}(a_s(Q^2))$ in Eq.(\ref{CSPp}) into     
the BLM scale of the effective charge  $\hat{a}_{Bjp}(Q^2)$ \cite{Rathsman}  
and to obtain N$^2$LO variant of  the results of Ref.\cite{Brodsky:1995tb}  
in the following form:    
 \begin{equation}
\label{BS2}
\bigg(1+\hat{a}_{D}(Q^2)\bigg)\bigg(1- \hat{a}_{Bjp}(Q^{*2}_{Bjp})\bigg)=1~.             
\end{equation}  
The  studies, analogous to those in \cite{Baikov:2010je}, should 
demonstrate  the N$^3$LO validity of this expression for the  second 
generalization of the Crewther relation.     

Within the  third variant of the generalized Crewther relation, 
proposed at N$^2$LO  in \cite{Kataev:2010dm} and considered   in 
detail at the N$^3$LO in \cite{Kataev:2010du},  the CSB-term in 
Eq.(\ref{CSPp})  is expressed through the following double expansion:
\begin{eqnarray}
\label{KM1}
 \Delta_{\rm{csb}}(a_s) = \sum_{n\geq 1}\bigg(\frac{\beta(a_s)}{a_s}\bigg)^n
{\cal P}_n(a_s) &=& \sum_{n\geq 1}\sum_{r\geq 1}\bigg(\frac{\beta(a_s)}{a_s}\bigg)^n P_n^{(r)}a_s^r \\
&=& \sum_{n\geq 1}\sum_{r\geq 1}\bigg(\frac{\beta(a_s)}{a_s}\bigg)^n P_n^{(r)} [{\rm{k,m}}] 
{\rm C_F^{\rm{k}}}{\rm  C_A^{\rm{m}}}~a_s^r , \label{KM2}
\end{eqnarray}
where the first  expansion parameter is the function $(\beta(a_s)/a_s)$ 
and the second expansion parameter is the coupling $a_s$ and 
${\rm C_F}$ and ${\rm C_A}$ are the quadratic Casimir operators of 
$SU(N_c)$ group.  
The indices ${\rm k,m}$ and $r$ in  Eq.(\ref{KM2}) are related as  
${\rm k+m}=r$ and the coefficients $P_n^{(r)}[{\rm{k,m}}]$ contain 
rational numbers and the  odd $\zeta$-functions. 
It should be stressed that the coefficients of  ${\cal P}_n(a_s)$ in  Eq.(\ref{KM1})
do not depend on $n_f$ which enter in the coefficients  $K_m$ of 
the first form of the generalized Crewther relation (\ref{BK}).  
However, 
contrary to the  first generalization of the Crewther relation, 
the validity of this  third generalization of 
the Crewther relation \cite{Kataev:2010du}   
is not yet   proved to all orders of perturbation theory. 
Another interesting problem is related to  the question  whether   
special features of   the    third extension of the  Crewther relation 
(\ref{KM1}) and Eq.(\ref{KM2}), namely, 
the $n_f$ independence of its coefficients, 
can be effectively used in practise. 
To analyze  this question in more detail, 
we supplement  the discussions of \cite{Kataev:2010du} by extra  
considerations  of the relations between 
the coefficients in the polynomials  ${\cal P}_n(a_s)$ 
in  Eq.(\ref{KM1}) and the ones in $D(a_s)$, 
$C_{Bjp}(a_s)$ and $\beta(a_s)$ functions.  
The $\beta$-expansion formalism \cite{MS04}
will be applied to this task and the results will be studied
in detail.

\section{Applications of $\beta$-expansion formalism}  
Consider  perturbative expansion of the normalized flavour non-singlet part 
of the   Adler function $D$ from Eq.(\ref{DA}) and the normalized $C^{Bjp}$ 
function from Eq.(\ref{Bjp}), namely, 
\begin{eqnarray}
\label{DAp}
D(a_s)&=&1 + \sum_{n=1} d_n~a_s^n~;~~
\\ \label{CBp}
C^{Bjp}(a_s)&=&1 + \sum_{l=1} c_l~a_s^l~.
\end{eqnarray}  
The QCD coupling constant $a_s$ obey the renormalization group equation with the 
QCD $\beta$-function which we will define as 
 \begin{equation}
\label{bf}
\mu^2 \frac{d}{d \mu^2} a_s=
\beta(a_s)=- a_s^2\left(\beta_0 + \beta_1 a_s + \beta_2 a_s^2 +\ldots\right).
\end{equation}  
 
Within the $\beta$-expansion formalism of Ref. \cite{MS04}, 
instead of commonly used representation of the coefficients of perturbative expansions for the
renormalization-group invariant quantities in  
powers of ${\rm T_F n_f}$ and the colour group factors,     
one should  
consider their expansions in powers of the    
$\beta_0$, $\beta_1$, $\beta_2 \ldots$ of the $\beta$-function.
For the quantities defined by the Eqs.(\ref{DAp}) and (\ref{CBp}) 
the coefficients of this expansion approach are $d_n[n_0,n_1,\ldots]$, $c_l[n_0,n_1,\ldots]$.
Their first argument corresponds to the term with $n_0$ powers of $\beta_0$, $\beta_0^{n_0}$, 
the second one -- $n_1$
 powers of $\beta_1$, $\beta_1^{n_1}$, and so on. 
The elements $d_n[0,0,\ldots,0]$, $c_l[0,0,\ldots,0]$  represent 
``genuine'' corrections
with powers $n_i=0$ of  all coefficients $\beta_i$. 
The latter elements  coincide with  expressions for the  
coefficients  $d_n$,  $c_l$ in the  imaginary   
case of the nullified QCD   $\beta$--function in 
all orders of perturbation theory is considered. 
This case  corresponds to restoration of CS of some quantum field model and will 
be considered here as a technical trick. 
If all  arguments $n_i$ after index $m$   of the elements
 $d_n[\ldots,m,0,\ldots,0]$ ($c_l[\ldots,m,0,\ldots,0]$)
are equal to zero, then, for the sake of a simplified notation, 
we omit these arguments and write these elements as $d_n[\ldots,m]$.
For the clarification of the $\beta$-expanded view of the coefficients in
perturbation series for physical quantities we consider the 
corresponding representations of the several
terms of Eq.(\ref{DAp}), namely, 
\begin{eqnarray}
\label{eq:d_2}
d_2&=&\! \beta_0\,d_2[1]
  + d_2[0]\, ,\\
\label{eq:d_3}  d_3
&=&\!
  \beta_0^2\,d_3[2]
  + \beta_1\,d_3[0,1]
  + \beta_0\,d_3[1]
  + d_3[0]\, , \\ 
  d_4
   &=&\! \beta_0^3\, d_4[3]
     + \beta_1\,\beta_0\,d_4[1,1]
     + \beta_2\, d_4[0,0,1]
     + \beta_0^2\,d_4[2]
     + \beta_1\,d_4[0,1]
     + \beta_0\,d_4[1]
     + d_4[0]\,,  \label{eq:d_4} \\
 d_{n}
   &=&\! \! \! \! \! \!~~\beta_0^{n-1}\! d_{n}[n\!-\!1]+ \ldots   
\label{eq:d_n} 
\end{eqnarray}
The same ordering in the $\beta$-function coefficients can be applied to the 
coefficients $c_l$.  
The expressions like Eq.(\ref{eq:d_2}--\ref{eq:d_n}) are unique. 
The first of them, Eq.(\ref{eq:d_2}), 
is the basis of the standard BLM prescription \cite{Brodsky:1982gc}.  
 
The coefficients  $d_n[n-1]$ are identical to the terms  generated 
by the renormalon chain insertions  and
 can be obtained  from the results of Ref.\cite{BK93}.
 The clarification of the physical and
mathematical origin of other elements is a separate and not straightforward task. 
The problem
of getting diagrammatic representation for different contributions into the $\beta$-expanded 
coefficients was considered in [39]. 
Further on in this Section we specify how to obtain $\beta$-expanded   results 
at the level of order $a_s^3$-corrections.

Together with Eq.(\ref{eq:d_2}--\ref{eq:d_n}), Eq.(\ref{KM1}) gives the possibility
to express the sum of the elements of  $n$--loop $\beta$-expanded coefficients  
through the ones which result from $(n-1)$--loop calculations.

We will use the property, that Eq.(\ref{Cr}) is satisfied in the CS  limit of QCD, namely, 
in the case when the $\beta$- function has  identically zero coefficients
$\beta_i=0$ for $i\geq 0$.  
In this model, the Crewther relation (\ref{Cr}) can be rewritten as
\begin{equation}  
\label{NC}
 D_0 \cdot C^{Bjp}_0 = 1,
\end{equation} 
where the expansions for the functions 
$D_0$ and $C^{Bjp}_0$, analogous to the ones of Eq.(\ref{DAp}), 
Eq.(\ref{CBp}), will contain the coefficients of  genuine content only,
namely,  $d_n~(c_n) \equiv d_n[0]~(c_n[0])$. 

Equation (\ref{NC}) provides evident relation between the genuine  elements in any loops, 
namely,  
\begin{equation}
c_n[0] + d_n[0] + \sum_{l=1}^{n-1} d_l[0] c_{n-l}[0]=0.
\label{eq:CI-PT0}
\end{equation}
In particular, they express the 
  yet unknown genuine parts  of the 
5-loop terms   $d_4, c_4$, 
through the 4-loop results already known from the analysis in \cite{MS04}:
\begin{equation} 
\label{eq:k4-d4}
c_4[0] + d_4[0]= 2d_1 d_3[0]-3d_1^2d_2[0]+(d_2[0])^2+d_1^4.
\end{equation} 
This equation contains contributions proportional to 
${\rm C_F}$ and ${\rm C_A}$.
Note that to check the perturbatively
quenched QED approximation for $d_4$ \cite{Baikov:2008jh}, 
available from  \cite{Baikov:2008cp},
it was suggested in \cite{Kat08} to use the relation that arises from the
projection of the relation (\ref{eq:k4-d4}) onto the maximum power of 
${\rm C_F}$, namely, 
${\rm C_F}^4$.

The $\beta$--expanded  form  for the   $d_3$-term  was obtained in \cite{MS04} 
by means of a careful consideration 
of the analytical ${\cal O}(a_s^3)$ expression for the  
Adler function $D(a_s,{\rm n_f,n_{\tilde{g}}})$ with the 
${\rm n_{\tilde{g}}}$ MSSM gluino  multiplets, 
obtained in  \cite{Chetyrkin:1996ez} 
\footnote{The 3-loop contribution 
of light gluinos coincide with the numerical 
result in \cite{Kataev:1983at}, while at the 4-loop  
analytical result for  gluino contribution, 
evaluated in \cite{Chetyrkin:1996ez},
was confirmed in \cite{Clavelli:1996zm}.}. 
The element $d_3[2]$,  which is proportional  
to the maximum power $\beta_0^2$ in (\ref{eq:d_3}), 
can be fixed in a straightforward way. 
Then one should separate the
contributions $\beta_1\, d_3[0,1]$ and 
$\beta_0\,d_3[1]$ to the $d_3$-term. 
They  both are linear in the  number of quark  flavours ${\rm n_f}$.
Their separation is possible if one uses additional degrees of freedom 
-- the gluino contributions mentioned above and  
labelled here by their ${\rm n_{\tilde{g}}}$ multiplet  number.
In  this way, one can get   the explicit form for the functions
${\rm n_f}={\rm n_f}(\beta_0,\beta_1)$ and ${\rm n_{\tilde{g}}}=
{\rm n_{\tilde{g}}}(\beta_0,\beta_1)$. 
They  can be obtained 
after taking into account the gluino contributions 
to the first two  coefficients of the  QCD  $\beta$-functions   
known from the  two-loop  calculations performed in 
\cite{Machacek:1983fi}.
Finally, one arrives at the    
expressions for the coefficients in Eqs.(\ref{eq:d_2}--\ref{eq:d_3}) 
presented in \cite{Kataev:2010du}, 
\begin{eqnarray}
d_1&=&\frac{3}{4}{\rm C_F};~
d_2[1]= \bigg(\frac{33}{8} - 3\zeta_3\bigg){\rm C_F};~
d_2[0]= -\frac{3}{32}{\rm C_F^2}  + \frac{1}{16}{\rm C_FC_A};~ 
\label{D-21} \\ \nonumber 
d_3[2]&=&\bigg(\frac{151}{6}-19\zeta_3\bigg){\rm C_F} \\ 
d_3[1]&=&
    \bigg(-\frac{27}{8} - \frac{39}{4}
    \zeta_3 + 15\zeta_5\bigg){\rm C_F^2}
-\bigg(\frac{9}{64} - 5\zeta_3 +\frac{5}{2}\zeta_5\bigg)
{\rm C_FC_A} \label{D-31}; \\
d_3[0,1]&=& \bigg(\frac{101}{16}-6\zeta_3\bigg){\rm C_F};~ 
d_3[0]= - \frac{69}{128}{\rm  C_F^3}+ \frac{71}{64}{\rm C_F^2C_A}+
    \bigg(\frac{523}{768}- \frac{27}{8}\zeta_3\bigg){\rm C_FC_A^2}~.  \label{D-30}
\end{eqnarray} 
which differ from the ones, originally obtained in Ref. \cite{MS04},
by the renormalization factor only. 
Let us emphasize that gluinos are used here as a pure technical device 
to reconstruct the $\beta$-function expansion of the perturbative coefficients. 

Using  the relation (\ref{eq:CI-PT0}) 
for  $n=2$ and $n=3$    and the already fixed $d_2[0]$ and $d_3[0]$-terms 
we get the expression  for the $c_2[0]$ and  
$c_3[0]$ coefficients. 
Their  knowledge allowed us to fix   other elements in the  
 $c_2[\ldots]$ and $c_3[\ldots]$ terms \cite{Kataev:2010du},    
without attracting additional gluino degrees of freedom. 
The results obtained in \cite{Kataev:2010du}  read:
\begin{eqnarray}
c_1&=&-\frac{3}{4}{\rm  C_F};~ 
c_2[1]= -\frac{3}{2}{\rm C_F};~
c_2[0]= \frac{21}{32}{\rm C_F^2}-\frac{1}{16}{\rm C_FC_A};  \\
c_3[2]&=&-\frac{115}{24}{\rm C_F};~
c_3[1]= \bigg(\frac{83}{24}- \zeta_3\bigg){\rm C_F^2} + 
\bigg(\frac{215}{192}- 6 \zeta_3+\frac{5}{2}\zeta_5\bigg){\rm C_FC_A};\label{c-31} \\ 
c_3[0,1]&=&\bigg(-\frac{59}{16}+3\zeta_3\bigg){\rm C_F};~
c_3[0] = - \frac{3}{128} {\rm C_F^3} -\frac{65}{64} {\rm C_F^2C_A}-
\bigg(   \frac{523}{768} - \frac{27}{8}\zeta_3\bigg){\rm C_FC_A^2}. 
\label{c-30}
\end{eqnarray}
Note that the  approximations  for the coefficients 
of the   Adler function and of  the  Bjorken polarized sum rule,  
 similar to those of Eqs.(\ref{eq:d_2}--\ref{eq:d_3}) but 
with the  terms proportional to the powers of $\beta_0$ only, 
were studied   in \cite{LovettTurner:1995ti}.
These expressions  from Ref.\cite{LovettTurner:1995ti} can be compared 
with the results of Eqs.(\ref{D-21}-\ref{c-30}).     
     
Consider now one of the applications of the $\beta$-expansion formalism \cite{MS04}. 
Substituting the $\beta$-expanded expressions for $d_{i}$ and $c_{i}$ into    
the general relations of Eq.(\ref{KM1}) 
we  found  the following identities \cite{Kataev:2010du}: 
\begin{eqnarray} 
\nonumber
{\cal P}_1(a_s)&=&a_s \bigg\{P^{(1)}_1+a_s P^{(2)}_1 + 
a_s^2 P^{(3)}_1 \bigg\} \\  \label{P1} 
&=&- a_s \bigg\{c_2[1] + d_2[1]+a_s\Big(c_3[1] + d_3[1]+ d_1\big(c_{2}[1]-d_{2}[1]\big) \Big) \nonumber  \\ 
&& + a_s^2 
\Big(c_4[1]+d_4[1]+d_1
\big(c_3[1]-d_3[1]\big)+d_2[0]c_2[1]+d_2[1]c_2[0]\Big)\bigg\} \\ 
\nonumber 
{\cal P}_2(a_s)&=&a_s \big\{P^{(1)}_2+a_sP_2^{(2)} \big\} \\
\label{P2}
&=&a_s\bigg\{c_3[2] + d_3[2]+a_s\Big( 
c_4[2]+d_4[2]
-d_1(c_3[2]-d_3[2])\Big)\bigg\}  \\
\label{P3}
{\cal P}_3(a_s)&=&a_s \bigg\{ P^{(1)}_3 \bigg\}  \nonumber \\
&=&
-a_s\big\{c_4[3] + d_4[3]\big\}= a_s{\rm C_F} \bigg(\frac{307}{2}-\frac{203}{2}\zeta_3-45\zeta_5\bigg)
\\  
\label{Pn}
{\cal P}_n(a_s)&=&
a_s  P^{(1)}_n =
(-1)^n a_s\big\{c_n[n-1] + d_n[n-1]\big\}  
\end{eqnarray} 
The elements $d_n[n-1]~(c_n[n-1])$ can be obtained 
from the results in \cite{BK93} for the  leading renormalon chain 
insertions.
The   elements  $d_n[l], (l< n-1)$  stem from  the 
subleading renormalon chains. 
Different relations between the elements $d_4~(d_n)$ and $c_4~(c_n)$ can 
be derived from
 the expression (\ref{KM1}) which is definitely true at 
the 5-loop level \cite{Kataev:2010du}. 
Its coefficients were obtained  by us analytically 
from  the requirement of their ${\rm n_f}$ independence \cite{Kataev:2010du}.
We also used the  property    
of universality of the weight function  ${\cal P}_1~({\cal P}_n)$ of   
the different $\beta_i$-terms. This property implies that  
the first  term  ${\cal P}_1(a_s)$ in Eq.(\ref{P1}) generates  
the following chain of equations: 
 \begin{eqnarray}
  \label{eq:k_n1-d_n1}
P^{(1)}_1&=& - c_2[1] - d_2[1] =- c_3[0,1] - d_3[0,1]= -c_4[0,0,1]-d_4[0,0,1] = \ldots \nonumber \\
&=&  -c_n[\underbrace{0,0,\ldots, 1}_{n-1}] - d_n[\underbrace{0,0,\ldots, 1}_{n-1}] =
 \left(-\frac{21}8+3\zeta_3 \right){\rm C_F}
 \end{eqnarray}
which  fixes the corresponding sums of $c_n$ and $d_n$ in any order by 
the universal first term $P^{(1)}_1$ in the polynomial ${\cal P}_1$. 
The second  term of ${\cal P}_1$ in Eq.(\ref{P1}) 
defines a similar chain of equations    
\begin{eqnarray} \label{k_401-d_401}
P^{(2)}_1 &=& -c_3[1] - d_3[1]- d_1\left(c_{2}[1] - d_{2}[1]\right)=
-c_4[0,1] - d_4[0,1]-d_1\left(c_3[0,1]-d_3[0,1]\right) = \ldots \nonumber \\
&=&-c_n[\underbrace{0,\ldots, 1}_{n-2}]-d_n[\underbrace{0,\ldots, 1}_{n-2}]- 
d_1\Big( c_{n-1}[\underbrace{0,\ldots, 1}_{n-2}]- d_{n-1}[\underbrace{0,\ldots, 1}_{n-2}]\Big)=
\ldots \nonumber \\
&=& 
\bigg(\frac{397}{96} + \frac{17}{2}\zeta_3 -15\zeta_5\bigg)
{\rm C_F^2}
-
 \bigg(\frac{47}{48}-\zeta_3 \bigg){\rm C_F C_A}  
\end{eqnarray}
where  the  explicit expression for $P^{(2)}_1$  is  already known  
(see \cite{Kataev:2010du}). 
The expressions 
for $P^{(3)}_1$ and  $P^{(2)}_2$  are  determined by  
Eqs.(\ref{P1}) 
and (\ref{P2}) respectively, and read 
\begin{eqnarray}
\label{P3cd}
P^{(3)}_1&=& -c_4[1]-d_4[1]-d_1
\big(c_3[1]-d_3[1]\big)+d_2[0]c_2[1]+d_2[1]c_2[0]  \\ \nonumber 
   \\ \label{P2cd} 
P^{(2)}_2&=&
c_4[2]+d_4[2]
-d_1(c_3[2]-d_3[2])
\end{eqnarray}
where their  concrete  form is   known from the studies 
of   Ref. \cite{Kataev:2010du}.

Thus, $P^{(3)}_1$ and  $P^{(2)}_2$ 
depend on still unknown contributions $d_4[1]~,~d_4[2]$ and  
$c_4[1]~,~c_4[2]$ for  the $\beta$-expanded form  of the
  general $SU(N_c)$-group  
expressions for  the   5-loop terms $d_4$ and $c_4$ obtained in \cite{Baikov:2010je}. 
Therefore, to reformulate  the  
considerations of \cite{Kataev:2010du} within the $\beta$-expansion approach,  
it is necessary to  determine these unknown contributions to $d_4$ and $c_4$.          
 
The theoretical problem mentioned above  may be solved  after extra  
analytical evaluation  of the  
gluino  contributions to the 5-loop perturbative coefficients   
$d_4$ and $c_4$  and applications  of  the  results of evaluation of the   
gluino contributions to the 3-loop   
QCD $\beta$-function \cite{Clavelli:1996pz}.
The knowledge of these inputs  from the $N=1$ SUSY model may  also allow a
better understanding of special features and constraints on  
the elements of the sum of $d_5+c_5$ coefficients   
which can be useful for study of the different forms of   
QCD generalizations of the Crewther relation in high orders of 
perturbation theory.

\section{The  constraints for the structure of the   5-loop 
analytical results} 
 
In this section, we explain how the relations 
discussed above and, in particular, the ones of Eqs.(\ref{D-21}-\ref{c-30})   
allow one to get an additional constraint which 
confirms the correctness of the  analytical 
results of Ref. \cite{Baikov:2010je}.

We  first  apply the $\beta$-expansion approach to the sum  $d_4+c_4$.
As the next step we consider the concrete  number of flavours which   
nullify the $\beta$-function coefficients $\beta_i$.  This is done by  
fixing  $n_i$ as a roots of the coefficient  $\beta_i(\rm {n_f}=n_i)$.
The condition  $\beta_0({\rm n_f}=n_0)=0$, which gives 
 ${\rm T_F} n_0=(11/4){\rm C_A}$, was studied by Banks--Zaks
some time ago \cite{Banks:1981nn}. Applying this ansatz to the sum of 
the $\beta$-expanded forms for the sum of the analysed  5-loop terms,  
we get \cite{Kataev:2010du}
\begin{equation} 
c_4(n_0) + d_4(n_0)\!=\!c_4[0]+d_4[0] + \beta_2(n_0)\left(c_4[0,0,1]+d_4[0,0,1]\right)
+ \beta_1(n_0)\left(c_4[0,1]+d_4[0,1]\right).
 \label{eq:check1}
 \end{equation}  
The terms in the r.h.s. of Eq.(\ref{eq:check1}) are  already known from the 
r.h.s. of Eq.(\ref{eq:k4-d4}) for ${\rm c_4[0]+d_4[0]}$,  
r.h.s. of Eq.(\ref{eq:k_n1-d_n1}) for the $P^{(1)}_1$-coefficient, and   
r.h.s. of Eq.(\ref{k_401-d_401}) for the $P^{(2)}_1$- term. Substituting 
the concrete analytical results into these expressions, we obtain 
\cite{Kataev:2010du}  
\begin{eqnarray} 
\label{BZ}
{\rm d_4}(n_0) + {\rm c_4}(n_0)&=& 
-\frac{333}{1024}{\rm  C_F^4}-{\rm C_A C_F^3} \left(\frac{1661}{3072}-
\frac{1309}{128}\zeta_3+
\frac{165 }{16}\zeta_5\right)  \nonumber \\
&&-{\rm C_A^2 C_F^2}  \left(\frac{3337}{1536}+\frac{7}{2} \zeta_3+
   \frac{105}{16} \zeta_5\right)-
  {\rm  C_A^3  C_F} \left(\frac{28931}{12288}-\frac{1351}{512}
   \zeta_3\right).~~~
\end{eqnarray}    
Then applying the Banks-Zaks way of fixation of $\rm{n_f}$  
to the concrete analytical expression for $c_4({\rm n_f})+d_4({\rm n_f})$, 
which follows from the work of  Ref.\cite{Baikov:2010je}, 
we reproduced the r.h.s. of  Eq.(\ref{BZ}) \cite{Kataev:2010du}. 
This agreement gave us an extra argument in favour of the correctness of the  
results of distinguished computer  analytical calculations of the
INR-Karlsruhe-SINP group  \cite{Baikov:2010je}.  
Moreover, having a look at the r.h.s. of Eq.(\ref{BZ}),
we observe the absence of the $\zeta_7$ and $\zeta_3^2$-terms which 
exist in analytical expressions of both $d_4$ and $c_4$ 
(see Ref.\cite{Baikov:2010je}). 
This feature observed in Ref. \cite{Kataev:2010du} 
confirms the foundation of the proportionality of two    transcendentalities mentioned above   
to the first coefficient $\beta_0$ of the QCD $\beta$-function   
\cite{Baikov:2010je}.

In the same way, taking ${\rm n_f}=n_1, 
(\beta_1(n_1)=0)$  one can get the following   constraint:  
\begin{eqnarray} 
c_4(n_1) + d_4(n_1)&=&c_4[0]+d_4[0] + \beta_0(n_1)\left(c_4[1]+d_4[1]\right)
 + \beta_0^2(n_1)\left(c_4[2]+d_4[2]\right) \nonumber \\
&& +\beta_2(n_0)\left(c_4[0,0,1]+d_4[0,0,1]\right)
+ \beta_0^3(n_1)\left(c_4[3]+d_4[3]\right).
~~ \label{eq:check2}
 \end{eqnarray}
The terms $c_4[0]+d_4[0]$ and $c_4[0,0,1]+d_4[0,0,1]$ 
in the r.h.s. of 
Eq. (\ref{eq:check2}) can be obtained like in the  previous case, while 
$c_4[3]+d_4[3]$ follows from Eq.(\ref{P3}). 
The term $c_4[1]+d_4[1]$ can be extracted from  $P^{(3)}_1$ in 
Eq.(\ref{P3cd}) and the  term $c_4[2]+d_4[2]$ 
can be extracted from $P^{(2)}_2$ 
in Eq.(\ref{P2cd}). The requirement 
$\beta_1(n_1)=0$ leads to 
the following  expression  for ${\rm n_f}$:  
\begin{equation}
{\rm n_f}=\frac{17{\rm C_A}^2}{(10{\rm C_A}+6{\rm C_F}){\rm T_F}}\ .
\end{equation}
For this expression we 
get a more complicated analytical structure of the r.h.s. 
of Eq.(\ref{eq:check2}) which does  not reveal  any special 
cancellations of the transcendental functions. In view of this 
we do not present it here  in the    explicit  form.  
\vspace{2mm}

{\bf Acknowledgements.}
One of us (ALK) is grateful to the members of the Organizing Committee of 
 XIXth International 
Workshop on High Energy Physics and Quantum Field Theory for invitation  
to present the plenary talk. We wish to thank K.G.Chetyrkin, E.W.N. Glover and 
 A.G. Grozin for constructive questions and comments.     
The work  of both of us was  supported by   
the RFBR, grant  No. 08-01-00686 and is 
carried out within the framework of the program of the RFBR, grant 11-01-00182. 
The work of ALK is also supported in part by grant NS-5525.2010.2.

{\bf  Appendix}

Throughout these studies we used the following expressions 
for the coefficients of the QCD $\beta$-function:
\begin{eqnarray}
\beta_0&=&\frac{11}{12}{\rm C_A}-\frac{1}{3}{\rm T_F n_f}, \label{beta0}\\
\beta_1&=& \frac{17}{24}{\rm C_A^2}-\frac{5}{12}{\rm C_A T_F n_f}
-\frac{1}{4}{\rm  C_F T_F n_f} \label{beta1} \\ 
\beta_2&=& \frac{2857}{3456}{\rm C_A^3}-\frac{1415}{1728}{\rm C_A^2T_Fn_f}-
\frac{205}{576}{\rm C_FC_AT_Fn_f} \nonumber \\
&&+\frac{79}{864}{\rm C_AT_F^2n_f^2}+\frac{1}{32}{\rm C_F^2T_Fn_f}
+\frac{11}{144}{\rm C_FT_F^2n_f^2}~\label{beta2}, 
\end{eqnarray}
They  are known from the 3-loop analytical calculations performed in  
\cite{Tarasov:1980au} and  confirmed later on in \cite{Larin:1993tp}.
In the case of  QCD,  
supplemented by ${\rm n_g}$ number of flavours of Majorano gluinos, 
the coefficients of the $\beta$-function receive additional 
$\overline{MS}$-scheme contributions obtained at 3-loops in \cite{Clavelli:1996pz},  
namely,
\begin{eqnarray}
\overline{\beta}_0&=&\beta_0-\frac{1}{6}{\rm C_An_g} \label{eq:beta-b0} \\
\overline{\beta}_1&=&\beta_1-\frac{5}{6}{\rm C_A^2n_g}-
\frac{1}{8}{\rm C_A^2n_g} \label{eq:beta-b1} \\
\overline{\beta}_2&=&\beta_2-\frac{247}{432}{\rm C_A^3n_g}
+\frac{7}{54}{\rm C_A^3T_Fn_fn_g}+\frac{11}{288}{\rm C_FC_A^2T_Fn_fn_g}
+\frac{145}{3456}{\rm C_A^3n_g^2}\label{eq:beta-b2}
\end{eqnarray}
Up to the 2-loop level they are scheme-independent. 
When  SUSY is not violated, the 
3-loop calculations should be performed in the dimensional reduction 
and $\overline{DR}$-scheme which preserve supersymmetry at the 
3-loop level.  
However, since  we are interested in the contributions of 
 ${\rm n_g}$ number 
of gluiono flavours only, we limit ourselves to the 
considerations of the part of the  $N=1$   SUSY 3-loop  contributions 
evaluated in the $\overline{MS}$-scheme.  

\end{document}